# Coupled cluster theory based on quantum electrodynamics: Physical aspects of closed shell and multi-reference open shell methods

Sambhu N. Datta*
Department of Chemistry, Indian Institute of Technology – Bombay, Powai, Mumbai 400 076, India
E-mail: sndatta@chem.iitb.ac.in

## Abstract

Electrodynamical coupled cluster (CC) methodologies have been formulated employing standard QED Hamiltonian that is written in Coulomb gauge while using the DF and the MCDF pictures of the matter field for closed-shell and open-shell cases respectively. The general methodology employs a radiative cluster, pure matter clusters and their pair modifications, and a number state distribution of photons in thermal equilibrium. The closed-shell treatment relies on the customary CC approach. For open shells, QED and electron correlation through CC are treated on the same footing. An averaging over the radiation state is done to generate Lamb, Breit and hyperfine interactions from the radiative cluster. Because of the thermal distribution, it leaves a residual transverse interaction that may modify the static correlation in open shells. Dynamical correlation effects are determined next by using the exponential matter cluster in traditional ways of single- and multi-reference CC. When the matter cluster is extended to include de-excitations to negative-energy levels, vacuum polarization effects are generated from the pair part of Coulomb interaction. The dynamical correlation energy includes relativistic corrections as well as QED contributions, namely, Lamb, Breit, hyperfine and pair energy.

This work has three novelties: (i) QED interactions (Lamb, Breit and hyperfine) are obtained from a single procedure based on the radiative cluster; (ii) pair energy is determined from an extended matter cluster formalism; and (iii) additional correlation energy can be had from radiative effects and pair terms, while the option for higher order pair energy in high-Z atoms is kept open. The open-shell formalism has one more novelty in finding an additional static correlation in certain cases when the radiation is not isotropic.

## 1. Introduction

Electrodynamical coupled cluster (CC) methodologies have been formulated [1] using
  i. *Standard QED Hamiltonian written in Coulomb gauge*;
 ii. Sucher [2] – Mittleman [3] "*Fuzzy*" picture of particles and antiparticles, namely, the Dirac-Fock (DF) picture of the matter field for closed shell cases, and the multi-configuration DF (MCDF) and multi-configuration self-consistent-field (MCSCF) pictures for open shell atoms or molecules;
iii. The *standard particle-antiparticle vacuum* that has been traditionally used by theoretical physicists starting from Feynman and Dyson;
   (The choice of the particle-hole vacuum, where the ground state configuration of "*N* electrons and *zero* positron" stands for vacuum, works in the nonrelativistic quantum chemistry without positrons, but it spoils the symmetry and equivalence of the particle world and the antiparticle world and introduces additional complications and algebraic requirements in a relativistic theory);

    iv.    A radiative cluster, pure matter clusters, and their pair modifications;
    v.    Number state distribution of photons in thermal equilibrium.

The methodology involves the usual CC approach for closed shells [4], and for open shells, QED and electron correlation are treated on the same footing [5].

The involved theory has the following operational characteristics.
a) Averaging over the radiation state is done to generate Lamb, Breit and hyperfine interactions from the radiative cluster.
b) Thermal distribution leaves a residual transverse interaction that may modify the static correlation in open shells.
c) Dynamical correlation effects can be obtained from the exponential involving the matter cluster in traditional ways.
d) Extending the matter cluster to include de-excitations to negative-energy levels gives rise to vacuum polarization effects from the pair part of Coulomb interaction.
e) Dynamical correlation energy includes relativistic corrections as well as QED effects such as Lamb shift, Breit interaction, hyperfine interaction and pair energy.

## 2. What is Quantum Electrodynamics (QED)?

QED is QFT (Quantum field theory) at a higher level.

In QFT only the matter field is quantized, and the matter-radiation interaction is semi-classically treated. It is also called Second Quantization as the operator for the number of particles is quantized. It comes in terms of the matter destruction and creation operators $a$, $a^\dagger$, etc. or the field annihilation and creation operators $\psi(r,t) = a(t)\varphi(r)$, $\psi^\dagger(r,t) = \varphi^\dagger(r)a^\dagger(t)$, etc. (summation convention). In QED, the photon field (number) is also quantized such that the matter-radiation interaction is quantum theoretical.

Nonrelativistic QED has matter field using nonrelativistic bases. As such by QED one means relativistic QED that has matter field prepared from relativistic bases, and the corresponding formalism involves matter, antimatter and pair components.

## 3. What is Coupled Cluster Methodology?

The intermediately normalized ground state wave function is prepared from the mean-field ground state configuration by using an exponential operator containing the cluster $T$ of excitation operators,

$$|\Phi_N^0\rangle = e^{\hat{T}} |\Psi_N^0\rangle \qquad (1)$$

s.t.   $\langle\Psi_N^0 | \Phi_N^0\rangle = \langle\Psi_N^0 | e^{\hat{T}} | \Psi_N^0\rangle = 1.$

Of course one finds the correlation energy from

$$(\hat{H} - E_N^0)|\Phi_N^0\rangle = E_{correl} |\Phi_N^0\rangle \qquad (2)$$

See references [6]-[7] for closed shells, [8] for open shells, [9]-[11] for open shell and general cases, [12]-[13] for Fock Space Pariser-Parr-Pople (PPP) multi-reference coupled cluster (MRCC), [14] for general Fock Space MRCC, [15] for molecular applications, and [16] for MRCC in Hilbert space.

**Varieties**

CC may contain only the doubles. Thus in CCD,
$$\hat{T} = \hat{T}_2 = \sum_{\substack{m<n \\ r<s}} C_{mn}^{rs} a_r^\dagger a_s^\dagger a_n a_m. \tag{3}$$

It is possible to expand the exponential and find quadruples as
$$|\Phi_N^0\rangle = (1 + \hat{T}_2 + \hat{T}_4 + ...)|\Psi_N^0\rangle \tag{4}$$

where
$$\hat{T}_4 = \sum_{\substack{m<m'<n<n' \\ r<s<t<u}} C_{mm'nn'}^{rstu} a_r^\dagger a_s^\dagger a_t^\dagger a_u^\dagger a_{n'} a_n a_{m'} a_m, \tag{5}$$

$C_{mm'nn'}^{rstu} = C_{mm'}^{rs} C_{nn'}^{tu} - \langle C_{mm'}^{rs} * C_{nn'}^{tu}\rangle$, etc.

The doubles cluster makes the most prominent contribution to correlation energy
$$\sum_{m<n, r<s} \langle \Psi_N^0 | \hat{H} | _{mn}^{rs}\rangle C_{mn}^{rs} = E_{correl}, \tag{6}$$

while $T_4$ improves the calculation by supplying quadruples and determines a better set of coefficients for the doubles from the relation
$$\langle _{mn}^{rs} | \hat{H} | \Psi_N^0\rangle + \sum_{p<q, t<u} \langle _{mn}^{rs} | \hat{H} - E_N^0 | _{pq}^{tu}\rangle C_{pq}^{tu} - \sum_{p<q, t<u} \langle \Psi_N^0 | \hat{H} | _{pq}^{tu}\rangle \langle C_{mn}^{rs} * C_{pq}^{tu}\rangle = 0. \tag{7}$$

These are the CCD equations.

The cluster $T = T_2$ can be modified by adding a cluster of singles, $T = T_1 + T_2$ where
$$\hat{T}_1 = \sum_{m,r} C_m^r a_r^\dagger a_m. \tag{8}$$

The resulting methodology is known as CCSD. It is computationally affordable, and it works better than MP2 and CISD. Of course the coefficients in $T_2$ become modified from those in CCD.

**Overall Assessment**

Nevertheless, CCSD is not accurate. It is found that
1) Triples need to be included in calculations near equilibrium geometry;
2) They are must in calculations near bond breaking;
3) They are also must for (MRCC) calculations on diradicals.
4) Quadruples are needed in calculations on double bond breaking.

CCSDTQ with selective slices of triples and quadruples can be described by
$$|\Phi_N^0\rangle = (1 + \hat{T}_1 + \hat{T}_2 + \hat{T}_3 + \hat{T}_4 + ...)|\Psi_N^0\rangle \tag{9}$$

where the higher order clusters can be written as
$$\hat{T}_3 = \sum_{\substack{m<m'<n \\ r<s<t}} C_{mm'n}^{rst} a_r^\dagger a_s^\dagger a_t^\dagger a_n a_{m'} a_m \tag{10}$$

etc. Coefficients at any order are to be partly determined by the lower-order coefficients.

## 4. The Mean field picture

Take $u_{m\sigma}$ ($v_{n\sigma}$) as the positive-energy (negative-mass) eigenspinors of the $N$-electron relativistic Fock operator

$$F = h_{D,ext} + \upsilon_{DF}(\mathbf{r}), \quad \upsilon_{DF}(\mathbf{r}) = \sum_{u_{m\sigma}}^{occ} [\mathrm{J}_{m\sigma}(\mathbf{r}) - \mathrm{K}_{m\sigma}(\mathbf{r})]. \tag{11}$$

The external-field Dirac operator is

$$h_{D,ext} = \beta mc^2 + c\boldsymbol{\alpha}\cdot\mathbf{p} + eA^0_{ext}(\mathbf{r}) \tag{12}$$

and the matter field is written as

$$\psi(\mathbf{r},t) = \psi_+(\mathbf{r},t) + \psi_-^\dagger(\mathbf{r},t) \tag{13}$$

where

$$\psi_+(\mathbf{r},t) = \underset{m\sigma}{\mathrm{S}}\, a_{m\sigma}(t) u_{m\sigma}(\mathbf{r}), \quad \psi_-(\mathbf{r},t) = \underset{p\sigma}{\mathrm{S}}\, b_{p\sigma}(t) v_{p\sigma}{}^*(\mathbf{r}) \tag{14}$$

in diagonal representation. Field operators follow equal time anti-commutation rule

$$[\psi_\pm(\mathbf{r},t),\, \psi^\dagger_\pm(\mathbf{r}',t)]_+ = \delta^3(\mathbf{r}-\mathbf{r}'), \quad [\psi_\pm(\mathbf{r},t),\, \psi_\pm(\mathbf{r}',t)]_+ = 0. \tag{15}$$

The equal time condition restricts covariance.

In the free particle picture, positive and negative energy solutions are distinctly known and the completeness relations hold separately for them. Sucher [17] showed that the non-perturbative use of the interaction associated with the Feynman gauge photon propagator in place of interaction associated with the Coulomb gauge propagator leads to energy levels that are "incorrect at the level of atomic fine structure". Basically the negative energy projector $\lambda_-(i) = \underset{p\sigma}{\mathrm{S}} |v_{p\sigma}(i)\rangle\langle v_{p\sigma}(i)|$ contributes to a trial spinor at order $(p/mc) \sim \alpha Z$ so that the energy levels can be incorrect at order $mc^2 \alpha^4 Z^4$.

### Different components of QED Hamiltonian

QED Hamiltonian components can be written in the field notation as shown below:

$$\hat{H}_{D,ext} = \int d^3r : \psi^\dagger(\mathbf{r}) h_{D,ext} \psi(\mathbf{r}) :, \tag{16}$$

$$\hat{H}_C = \frac{e^2}{2} \int d^3r_1 \int d^3r_2\, \hat{\rho}_D(\mathbf{r}_1) \frac{1}{|\mathbf{r}_1 - \mathbf{r}_2|} \hat{\rho}_D(\mathbf{r}_2), \tag{17}$$

$$\hat{H}_{QFT} = \hat{H}_{D,ext} + \hat{H}_C \Leftrightarrow H_{DC} \text{ (in coordinate representation)}. \tag{18}$$

The 4-current operators are found as

$$\hat{\rho}_D(\mathbf{r}) = :\psi^\dagger(\mathbf{r})\psi(\mathbf{r}): \quad \text{(Matter density operator)}$$
$$= \psi^\dagger_+(\mathbf{r})\psi_+(\mathbf{r}) + \psi^\dagger_+(\mathbf{r})\psi^\dagger_-(\mathbf{r}) + \psi_-(\mathbf{r})\psi_+(\mathbf{r}) + :\psi_-(\mathbf{r})\psi^\dagger_-(\mathbf{r}): \tag{19}$$

and

$$\mathbf{J}_D(\mathbf{r}) = c:\psi^\dagger(\mathbf{r})\boldsymbol{\alpha}\psi(\mathbf{r}): \quad \text{(Probability current operator)} \tag{20}$$

## 5. Hamiltonian operators

State vectors in the configuration space $|\Psi^n_N\rangle_{cs}$ satisfy

$$\Lambda_+ |\Psi_N^n\rangle_{cs} = |\Psi_N^n\rangle_{cs}, \tag{21}$$

where

$$\Lambda_+ = \prod_{i=1}^{N} \lambda_+(i), \quad \text{and} \quad \lambda_+(i) = \underset{m\sigma}{S} |u_{m\sigma}(i)\rangle\langle u_{m\sigma}(i)|. \tag{22}$$

The projected interaction is

$$\Lambda_+ \sum_{1 \leq i < j \leq N} \frac{e^2}{|\mathbf{r}_i - \mathbf{r}_j|} \Lambda_+ = \mathbf{H}_{C++} \Leftrightarrow \hat{H}_{C++} \tag{23}$$

The second-quantized operator $\hat{H}_{C++}$ is obtained by using only the first component of density in $\hat{H}_C$. In a similar way, the projected external Dirac operator is

$$\Lambda_+ \sum_{i=1}^{N} h_{D,ext}(i) \Lambda_+ = \mathbf{H}_{D,ext+} \tag{24}$$

corresponding to the operator $\hat{H}_{D,ext+}$ of QED.

The projected Hamiltonian of QFT is indeed the no-pair Hamiltonian restricted to the N-electron sector of Fock space [2]:

$$\hat{H}_{QFT}(projected) = \hat{H}_{D,ext+} + \hat{H}_{C++} \Leftrightarrow \hat{H}_{QFT}^{no-pair} = \hat{H}_{D,ext}^{no-pair} + \hat{H}_{C}^{no-pair}. \tag{25}$$

The no-pair operators are obtained using only the first and fourth components of the density operator. The corresponding Dirac-Fock energy of the ground state configuration is

$$E_N^0 = \langle \Psi_N^0 | \hat{H}_{QFT}(projected) | \Psi_N^0 \rangle = \langle \Psi_N^0 | \hat{H}_{D,ext+} | \Psi_N^0 \rangle + \langle \Psi_N^0 | \hat{H}_{C++} | \Psi_N^0 \rangle. \tag{26}$$

The operator $\hat{H}_C^{Pair} = \hat{H}_C - \hat{H}_C^{no-pair}$ does not contribute to the Dirac-Fock ground state configuration energy.

**The DF Hamiltonian (coordinate and field representations)**

The Dirac-Fock Hamiltonian is written as

$$\mathbf{H}_{DF} = \sum_{i=1}^{N} F(i) \Leftrightarrow \hat{H}_{DF} = \hat{H}_{D,ext} + \hat{V}_{DF}, \quad \hat{V}_{DF} = \int d^3r : \psi^\dagger(\mathbf{r}) \upsilon_{DF}(\mathbf{r}) \psi(\mathbf{r}) : \tag{27}$$

such that

$$\hat{H}_{DF} |\Psi_N^n\rangle = E_{DF}^n |\Psi_N^n\rangle. \tag{28}$$

It is obvious that $E_{DF}^0$ differs from $E_N^0$ by an additional average of the projected interaction energy.

## 6. Quantized radiation field

Photon creation and annihilation operators characterized by wave vector $\mathbf{k}$ and polarization vector $\lambda$ satisfy commutation rules

$$[A_{k\lambda}, A_{k'\lambda'}]_- = 0 \text{ and } [A_{k\lambda}, A_{k'\lambda'}^\dagger]_- = \delta_{k,k'} \delta_{\lambda,\lambda'}. \tag{29}$$

Also notice $\omega k = ck$, and $\hat{N}_{k\lambda}$ is the number operator $\hat{N}_{k\lambda} = A_{k\lambda}^\dagger A_{k\lambda}$. The number state is written as $|N_{k\lambda}\rangle$ s.t. $\langle N_{k\lambda} | N_{k'\lambda'}\rangle = \delta_{k,k'} \delta_{\lambda,\lambda'}$. One finds

$$A_{k\lambda}|N_{k'\lambda'}\rangle = \delta_{k,k'}\delta_{\lambda,\lambda'}(2\pi\hbar c^2/\omega_k)^{1/2}(N_{k\lambda})^{1/2}|N_{k\lambda}-1\rangle,$$
$$A_{k\lambda}^{\dagger}|N_{k'\lambda'}\rangle = \delta_{k,k'}\delta_{\lambda,\lambda'}(2\pi\hbar c^2/\omega_k)^{1/2}(N_{k\lambda}+1)^{1/2}|N_{k\lambda}+1\rangle, \quad (30)$$

that gives
$$\langle \hat{N}_{k\lambda}\rangle = (2\pi\hbar c/k)N_{k\lambda}. \quad (31)$$

Electromagnetic 4-potentials in Transverse gauge are given by
$$A^0(\mathbf{r},t) = \int d^3r' \frac{\hat{\rho}_D^0(\mathbf{r}',t)}{|\mathbf{r}-\mathbf{r}'|}, \quad \mathbf{A}(\mathbf{r},t) = \frac{1}{\sqrt{\Omega}}\sum_{k\lambda}\left[A_{k\lambda}\lambda e^{i(\mathbf{k}\cdot\mathbf{r}-\omega_k t)} + h.c.\right]. \quad (32)$$

where $\Omega$ is the volume in which the photons are counted. The Hamiltonian for unit volume is
$$\hat{H}_{rad}^0 = \Omega^{-1}\sum_{k\lambda}(k/2\pi\hbar c)\hat{N}_{k\lambda}\hbar\omega_k \quad (33)$$

so that by virtue of (31) the state vector $|\Psi_{rad}^0\rangle = |\{N_{k\lambda}\}\rangle$ offers the energy density
$$E_{rad}^0 = \Omega^{-1}\sum_{k\lambda}N_{k\lambda}\hbar\omega_k.$$

### Distribution of number states

Consider the general distribution
$$|k\lambda\rangle = \sum_N g_{Nk\lambda}|N_{k\lambda}\rangle \quad (34)$$

where $\sum_N g_{Nk\lambda}^2 = 1$. This indeed gives
$$\langle N_{k\lambda}\rangle = \frac{2\pi\hbar c}{k}\bar{N}_{k\lambda}, \quad \bar{N}_{k\lambda} = \sum_N g_{Nk\lambda}^2 N \quad (35)$$

such that
$$E_{rad}^0 = \Omega^{-1}\sum_{k\lambda}\bar{N}_{k\lambda}\hbar\omega_k. \quad (36)$$

For photons in thermal equilibrium,
$$g_{Nk\lambda} = \left(1-e^{-\hbar\omega_k/\tau}\right)^{1/2} e^{-N\hbar\omega_k/2\tau} \Leftrightarrow \sum_{N=0}^{\infty} g_{Nk\lambda}^2 = 1 \text{ for each } |k\lambda\rangle. \quad (37)$$

This gives
$$\bar{N}_{k\lambda} = (e^{\hbar\omega_k/\tau}-1)^{-1}, \quad E_{k\lambda} = \hbar ck\bar{N}_{k\lambda} = \hbar ck(e^{\hbar\omega_k/\tau}-1)^{-1} \text{ and } E_{rad}^0 = \Omega^{-1}\sum_{k\lambda}E_{k\lambda}. \quad (38)$$

### Interaction of radiation with matter

The covariant interaction between matter 4-current and EM 4-potential is given by
$$\hat{H}_{int} = e\int d^3r\, J_\mu A^\mu = \hat{H}_{int}^{(1)} + \hat{H}_{int}^{(2)}. \quad (39)$$

The longitudinal photon interaction is
$$\hat{H}_{int}^{(1)} = e^2\int d^3r\, \hat{\rho}_D(\mathbf{r},t)A^0(\mathbf{r},t), \quad (40)$$

as in Coulomb gauge,
$$eA^0(\mathbf{r},t) = \int d^3r' \frac{e\hat{\rho}_D(\mathbf{r}',t)}{|\mathbf{r}-\mathbf{r}'|}. \quad (41)$$

For the transverse photon interaction,
$$\hat{H}_{int}^{(2)} = -\frac{e}{c}\int d^3r\, \hat{\mathbf{J}}_D(\mathbf{r},t)\cdot\mathbf{A}(\mathbf{r},t). \quad (42)$$

Every pairwise interaction is counted twice in $\hat{H}_{int}^{(1)}$. One needs to keep only $\hat{H}_C = \hat{H}_{int}^{(1)}/2$ which is a further departure from the approximate covariance of the equal time representation [2].

Vector interactions in $\hat{H}_{int}^{(2)}$ are responsible for known QED effects such as
- Breit interaction of order $mc^2\alpha^4Z^4$ (electron-electron magnetic interaction plus retarded interaction)
- Lamb shift of order $mc^2\alpha^5Z^4$ (mainly a part of the electron self energy)
- Hyperfine interaction of order $(m^2/M)c^2\alpha^4Z^3$ (electron-nucleus magnetic interaction)

Hence the QED Hamiltonian is
$$\hat{H}_{QED} = \hat{H}_{rad}^0 + \hat{H}_{D,ext} + \hat{H}_C + \hat{H}_{int}^{(2)}. \tag{43}$$

## 7. Presently practiced relativistic CC

The pros and cons of the presently practiced relativistic CC calculations are outlined here. (1) In place of Schrödinger Hamiltonian and basis functions, choose Dirac Hamiltonian and spinors. (2) Replace integrals. (3) Do a straight-forward CC. (4) Atomic spinor orbitals and integrals are available free of cost. It is not surprising that most of the relativistic CC calculations have been done on atoms and atomic ions.

**Points of interest**

The CC is based on DF orbitals determined from either the projected DC Hamiltonian [18] or the projected DCB Hamiltonian [19]. Breit interaction energy is obtained as expectation value over DF ground state wave function. In the second procedure, it also contributes to the determination of the ground state configuration.

Error due to variation collapse is avoided at the DF level by using the approximate methods like matrix representation of operators [20-21] or the better approach of implementing the kinetic balance for lighter elements [22-25].

For finite basis, spurious negative energy spinors [26] are discarded to avoid effects like continuum dissolution so that the interaction is projected by default [3].

Numerical DF orbitals as basis sets can give an approximately correct projection [27].

MRCC treatment has also been formulated

Some authors base the relativistic CC on Douglas-Kroll-Hess transformation and use the two-component spinors in order to bypass problems with Breit interaction and variation collapse [28]. But calculation remains approximate through any finite order and requires a large basis set, with slow convergence and tedious evaluation of radiative effects. The errors remain inherent.

Often the spin-orbit splitting of basis orbitals is neglected. A relativistic treatment is for numerical accuracy that costs computational time, space and effort, and the effects of the spin-orbit interaction should be fully retained in selection of basis spinors [29].

**Summary and shortcomings**

Otherwise, it is doing nonrelativistic-type CC. This is a straight-forward work that has been termed by Sucher as "follow-your-nose" (FYN) [2]. What is left out in FYN? The imminent answer is the negative-energy (antiparticle) effects.

To the FYN CC, one may add chosen QED corrections of the lowest order(s). Then one gets straight-forward QED CC. What one loses are 1) formalistic maneuver and 2) vacuum polarization effects in detail.

**Conclusions**

It is logical to use the Hamiltonian of QED rather than starting at the halfway mark. The first objective is to obtain the QED interactions (Lamb, Breit and hyperfine) from a single procedure based on a radiative cluster. The second one is to implement these in the CC with matter, antimatter and photon fields.

## 8. QED CC

We write the product state vector as
$$|\Psi_0\rangle = |\Psi_N^0\rangle|\Psi_{rad}^0\rangle \qquad (44)$$
And the ground state energy values
$$E_N^0 = \langle\Psi_0|\hat{H}_{D,ext} + \hat{H}_{C++}|\Psi_0\rangle,$$
$$E_{rad}^0 = \langle\Psi_0|\hat{H}_{rad}^0|\Psi_0\rangle. \qquad (45)$$
We require *at least two* clusters, one for the radiative effect originating from $\hat{H}_{int}^{(2)}$ and the other for the matter correlation from $\hat{H}_{int}^{(1)}$.

**The Radiative Cluster**

One may notice:
The transverse interaction $\hat{H}_{int}^{(2)}$ linearly varies with $A$ and $A^\dagger$.
Its expectation value over $|\Psi_{rad}^0\rangle$ is zero.
It is bilinear in matter field, and because of Dirac α matrix, accommodates single-particle excitations.
The radiative cluster $\hat{T}_{int}^{(2)}$ is to work with $\hat{H}_{int}^{(2)}$ at least as a linear factor, and the product must give nonzero average values over radiation as well as matter states.
The simplest cluster is to be formed from singles and to linearly vary with $A$ and $A^\dagger$. Hence it is to differ from the terms in $\hat{H}_{int}^{(2)}$ only by a multiplicative factor for each intermediate state.

RSPT gives the second order correction $\langle\hat{V}\hat{Q}\hat{V}\rangle$ where
$$\hat{Q} = \underset{I}{S}|I\rangle\langle I|(E-\hat{H}_0)^{-1}|I\rangle\langle I|. \qquad (46)$$
The operator product $\hat{H}_{int}^{(2)}\hat{T}_{int}^{(2)}$ $(=\hat{V}\hat{Q}\hat{V})$ is to be manifestly hermitean, identifying $\hat{T}_{int}^{(2)} \simeq \hat{Q}\hat{V}$ as a possible cluster. Cluster $\hat{Q}\hat{V}$ creates differences at higher orders. At order 3, RSPT gives $\langle\hat{V}\hat{Q}\hat{V}\hat{Q}\hat{V}\rangle - \langle\hat{V}\rangle\langle\hat{V}\hat{Q}\hat{Q}\hat{V}\rangle$ while the exponential operator gives $\langle\hat{V}\hat{Q}\hat{V}\hat{Q}\hat{V}\rangle/2$ at second order. When the higher order terms become zero or negligibly small, operator $\hat{Q}\hat{V}$ suffices. Hence the lowest-order radiative cluster is written as

$$\hat{T}_{int}^{(2)} = \hat{Q}\hat{V} = \sum_{n}\sum_{k\lambda}\sum_{N'_{k\lambda}} |\Psi_N^n\rangle |N'_{k\lambda}\rangle [E_{DF}^0 - E_{DF}^n + (N_{k\lambda} - N'_{k\lambda})\hbar ck]^{-1} \langle N'_{k\lambda}|\langle \Psi_N^n| \times$$
$$\left(-\frac{e}{c\sqrt{\Omega}}\right) \int d^3r\, \hat{J}_D(r) \cdot \left[A_{k\lambda}\lambda e^{i(k\cdot r - \omega_k t)} + A_{k\lambda}^\dagger \lambda^* e^{-i(k\cdot r - \omega_k t)}\right]. \quad (47)$$

**Matter clusters**

The double excitation operator $\hat{T}_2$ is a staple in both nonrelativistic and relativistic CC treatments. It can be fortified by adding the singles, triples, quadruples, etc. The *first intermediately normalized state* is

$$|\tilde{\Phi}_0\rangle = e^{\hat{T}_{mat}}|\Psi_0\rangle = (1 + \hat{T}_1 + \hat{T}_2 + \hat{T}_3 + \hat{T}_4 + ...)|\Psi_0\rangle. \quad (48)$$

The clusters $\hat{T}_1$, $\hat{T}_2$, $\hat{T}_3$, and $\hat{T}_4$ are neither hermitean nor anti-hermitean.

The net cluster in the exponential is $(\hat{T}_{int}^{(2)} + \hat{T}_{mat})$. The *final intermediately normalized state* is

$$|\Phi_0\rangle = \frac{e^{\hat{T}_{int}^{(2)}}}{\langle e^{\hat{T}_{int}^{(2)}}\rangle_{\Psi_{rad}^0}}|\tilde{\Phi}_0\rangle = \frac{e^{\hat{T}_{int}^{(2)}}}{\langle e^{\hat{T}_{int}^{(2)}}\rangle_{\Psi_{rad}^0}} e^{\hat{T}_{mat}}|\Psi_0\rangle. \quad (49)$$

s. t. $\langle \Psi_0 | \Phi_0 \rangle = 1$.

.

## 9. Calculations for both Closed and Open Shells

Start with calculations on the radiation average:

$$\frac{\langle \hat{H}_{QED} e^{\hat{T}_{int}^{(2)}}\rangle_{\Psi_{rad}^0}}{\langle e^{\hat{T}_{int}^{(2)}}\rangle_{\Psi_{rad}^0}} \quad (50)$$

$$= E_{rad}^0 + \hat{H}_{D,ext} + \hat{H}_C + \langle \hat{H}_{int}^{(2)}\rangle_{rad}(1 - \langle \hat{T}_{int}^{(2)}\rangle_{rad}) + \langle \hat{H}_{int}^{(2)} \hat{T}_{int}^{(2)}\rangle_{rad} + O(mc^2\alpha^5 Z^5).$$

The operator $\hat{T}_{int}^{(2)}$ is of order $\alpha Z$, while the interaction $\hat{H}_{int}^{(2)} = -\frac{e}{c}\int d^3r\, \hat{J}_D(r,t)\cdot A(r,t)$ is of order $mc^2\alpha^3 Z^3$ and their product $\hat{H}_{int}^{(2)} \hat{T}_{int}^{(2)}$ is of order $mc^2\alpha^4 Z^4$. Thus in equation (50), $\langle \hat{H}_{int}^{(2)}\rangle_{rad}$, $\langle \hat{T}_{int}^{(2)}\rangle_{rad}$ and hence $\langle \hat{H}_{int}^{(2)}\rangle_{rad}\langle \hat{T}_{int}^{(2)}\rangle_{rad}$ all equal zero, and the second order energy correction $\langle \hat{H}_{int}^{(2)} \hat{T}_{int}^{(2)}\rangle_{rad}$ gives the electron self energy and Breit energy. The fourth order correction would be negligibly small for $\alpha^4 Z^4 \ll 1$. The following calculations will be useful.

$$\langle \Psi_{rad}^0 | \hat{H}_{int}^{(2)} | \Psi_{rad}^0 \rangle = -\frac{e}{c} \int d^3r \, \hat{\boldsymbol{J}}_D(\boldsymbol{r},t) \cdot \langle \Psi_{rad}^0 | \boldsymbol{A}(\boldsymbol{r},t) | \Psi_{rad}^0 \rangle,$$

$$\langle \Phi_\rho | \langle \Psi_{rad}^0 | \hat{H}_{int}^{(2)} | \Psi_{rad}^0 \rangle | \Phi_\sigma \rangle = -\frac{e}{c} \int d^3r \, \langle \Phi_\rho | \hat{\boldsymbol{J}}_D(\boldsymbol{r},t) | \Phi_\sigma \rangle \cdot \langle \Psi_{rad}^0 | \boldsymbol{A}(\boldsymbol{r},t) | \Psi_{rad}^0 \rangle,$$

$$\langle \Psi_{rad}^0 | \boldsymbol{A}(\boldsymbol{r},t) | \Psi_{rad}^0 \rangle = \frac{1}{\sqrt{\Omega}} \sum_k \left( \frac{2\pi\hbar c}{k} \right)^{1/2} \sum_\lambda \Big[ \, \lambda e^{i(\boldsymbol{k}\cdot\boldsymbol{r}-\omega_k t)} \sum_n (n+1)^{1/2} g_{nk\lambda} g_{(n+1)k\lambda}$$
$$+ \lambda^* e^{-i(\boldsymbol{k}\cdot\boldsymbol{r}-\omega_k t)} \sum_n n^{1/2} g_{nk\lambda} g_{(n-1)k\lambda} \, \Big].$$

$$\langle \Psi_{rad}^0 | \hat{T}_{int}^{(2)} | \Psi_{rad}^0 \rangle = 0.$$

$$\langle \Psi_{rad}^0 | \hat{H}_{int}^{(2)} \hat{T}_{int}^{(2)} | \Psi_{rad}^0 \rangle = \frac{e^2}{c^2 \Omega} \int d^3r \int d^3r' \, \hat{\boldsymbol{J}}_D(\boldsymbol{r},t) | \Psi_N^n \rangle \langle \Psi_N^n | \hat{\boldsymbol{J}}_D(\boldsymbol{r}',t)$$
$$\times \sum_{k\lambda} \frac{2\pi\hbar c}{k} \left[ \sum_N (E_{DF}^0 - E_{DF}^n + \hbar ck)^{-1} \left\{ (N+1)^{1/2}(N+2)^{1/2} g_{Nk\lambda} g_{(N+2)k\lambda} e^{i\boldsymbol{k}\cdot(\boldsymbol{r}+\boldsymbol{r}')} + (N+1) g_{Nk\lambda}^2 e^{i\boldsymbol{k}\cdot(\boldsymbol{r}-\boldsymbol{r}')} \right\} \right.$$
$$\left. + (E_{DF}^0 - E_{DF}^n - \hbar ck)^{-1} \left\{ N g_{Nk\lambda}^2 e^{-i\boldsymbol{k}\cdot(\boldsymbol{r}-\boldsymbol{r}')} + N^{1/2}(N-2)^{1/2} g_{Nk\lambda} g_{(N-2)k\lambda} e^{-i\boldsymbol{k}\cdot(\boldsymbol{r}+\boldsymbol{r}')} \right\} \right].$$
(51)

**To be more explicit**

Averaging over the photon states gives the second order energy correction due to $\hat{H}_{int}^{(2)}$ — the fourth order correction is negligibly small for $\alpha^4 Z^4 \ll 1$. In essence one obtains the effect of the interaction of one or two electron(s) absorbing (emitting) and subsequently emitting (absorbing) a transverse virtual photon. Use the transformation

$$\Omega^{-1} \sum_k \;\rightarrow\; \frac{1}{8\pi^3} \int d^3k \tag{52}$$

to get
$$\langle \Psi_{rad}^0 | H_{int}^{(2)} \hat{T}_{1,int}^{(2)} | \Psi_{rad}^0 \rangle \doteq \hat{H}_{SE}^{ext} + \hat{H}_{Breit}. \tag{53}$$

**Lamb Shift**

Most of the electron self-energy $\hat{H}_{SE}^{ext}$ in presence of the external potential can be discarded by renormalization, keeping only the part that exists even in radiation vacuum

$$\hat{H}_{SE,vac}^{ext} = \frac{e^2 \hbar}{4\pi^2 c} \int \frac{d^3k}{k} \sum_\lambda \int d^3r \sum_n \hat{\boldsymbol{J}}_D(\boldsymbol{r}) \cdot \lambda \, \frac{|\Psi_N^n\rangle\langle\Psi_N^n|}{E_{DF}^0 - E_{DF}^n - \hbar ck} \hat{\boldsymbol{J}}_D(\boldsymbol{r}) \cdot \lambda \tag{54}$$

Averaging over the transverse polarization gives an integral that linearly diverges [30]. When it is translated into the nonrelativistic limit, one sees that a renormalization of mass is necessary. The same is done by subtracting a similar correction for the free electron. Thus the visible part of this self-energy is obtained, but it is still logarithmically divergent. The divergence is finally removed by using the cut-off $k_{co}=mc/\hbar$ as the upper boundary of the $k$-integral and one has the effective Hamiltonian operator for Lamb shift. A general expression valid for the $n$th arbitrary DF state can be written using $mc^2 \gg |E_{DF}^{n'} - E_{DF}^n|$ and the realization that the involved $k$-integral is in reality a principal value integral:

$$\hat{H}_{Lamb} = \frac{2\alpha}{3\pi c^2} \int d^3r \sum_{n'(\neq n)} \hat{J}_D(r) |\Psi_N^{n'}\rangle \cdot \langle \Psi_N^{n'}| \hat{J}_D(r) \times (E_{DF}^{n'} - E_{DF}^n) \ln\left(\frac{mc^2}{|E_{DF}^{n'} - E_{DF}^n|}\right) \quad (55)$$

The self-energy interpretation accounts for 1040 Mc [30], more than 98% of $2s_{1/2} - 2p_{1/2}$ shift in hydrogen atom (1057.8 MHz) [31]. Lamb shift can also arise from vacuum polarization (–27 MHz), vertex corrections and higher order corrections [30, 32]. The calculated net shifts for H, D and He$^+$ differ from the experimental ones only in the sixth significant digit.

### Breit Interaction

For the remaining term in (53), the virtual photon energy is much greater than the excitation energy, and the sum over *N*-electron states can be replaced by unit operator, leading to the Breit operator

$$\hat{H}_{Breit} = -\frac{e^2}{4\pi^2 c^2} \int \frac{d^3k}{k^2} \sum_\lambda \Big[ (N_{k\lambda}+1) \int d^3r \int d^3r' e^{i\boldsymbol{k}\cdot(\boldsymbol{r}-\boldsymbol{r}')} \hat{J}_D(r) \cdot \lambda \hat{J}_D(r') \cdot \lambda$$
$$- N_{k\lambda} \int d^3r \int d^3r' e^{-i\boldsymbol{k}\cdot(\boldsymbol{r}-\boldsymbol{r}')} \hat{J}_D(r) \cdot \lambda \hat{J}_D(r') \cdot \lambda \Big]. \quad (56)$$

Summation over polarization vectors gives a sum of magnetic and retardation components of interactions,

$$\hat{H}_{Breit} = \hat{H}_{Magnetic} + \hat{H}_{Retarded},$$

$$\hat{H}_{Magnetic} = -\frac{e^2}{4\pi^2 c^2} \int \frac{d^3k}{k^2} \int d^3r \int d^3r' \, e^{i\boldsymbol{k}\cdot(\boldsymbol{r}-\boldsymbol{r}')} \hat{J}_D(r) \cdot \hat{J}_D(r'), \quad (57)$$

$$\hat{H}_{Retarded} = +\frac{e^2}{4\pi^2 c^2} \int \frac{d^3k}{k^2} \int d^3r \int d^3r' \, e^{i\boldsymbol{k}\cdot(\boldsymbol{r}-\boldsymbol{r}')} \frac{\hat{J}_D(r) \cdot k \hat{J}_D(r') \cdot k}{k^2}.$$

Singer transformation simplifies the magnetic part into the 1/r form

$$\hat{H}_{Magnetic} = -\frac{e^2}{2c^2} \int d^3r \int d^3r' \frac{\hat{J}_D(r) \cdot \hat{J}_D(r')}{|\boldsymbol{r}-\boldsymbol{r}'|} \quad (58)$$

while an identity

$$\frac{1}{2\pi^2} \int \frac{d^3k}{k^2} e^{i\boldsymbol{k}\cdot\boldsymbol{r}} \frac{\boldsymbol{a}\cdot\boldsymbol{k}\,\boldsymbol{b}\cdot\boldsymbol{k}}{k^2} = \frac{1}{2r}\left[\boldsymbol{a}\cdot\boldsymbol{b} - \frac{\boldsymbol{a}\cdot\boldsymbol{r}\,\boldsymbol{b}\cdot\boldsymbol{r}}{r^2}\right] \quad (59)$$

is used to convert the retardation part into the Casimir form

$$\hat{H}_{Retarded} = \frac{e^2}{4c^2} \int d^3r \int d^3r' \frac{1}{|\boldsymbol{r}-\boldsymbol{r}'|}\left[\hat{J}_D(r)\cdot\hat{J}_D(r') - \frac{\hat{J}_D(r)\cdot(\boldsymbol{r}-\boldsymbol{r}')\hat{J}_D(r')\cdot(\boldsymbol{r}-\boldsymbol{r}')}{|\boldsymbol{r}-\boldsymbol{r}'|^2}\right]. \quad (60)$$

One gets

$$\hat{H}_{Breit} = -\frac{e^2}{4c^2} \int d^3r \int d^3r' \frac{1}{|\boldsymbol{r}-\boldsymbol{r}'|}\left[\hat{J}_D(r)\cdot\hat{J}_D(r') + \frac{\hat{J}_D(r)\cdot(\boldsymbol{r}-\boldsymbol{r}')\hat{J}_D(r')\cdot(\boldsymbol{r}-\boldsymbol{r}')}{|\boldsymbol{r}-\boldsymbol{r}'|^2}\right]. \quad (61)$$

After the self-energy corrections, one finds the usual form in coordinate representation,

$$\mathbf{H}_{Breit} = \sum_{1 \leq i < j \leq N} B(i,j) \quad (\Leftrightarrow \hat{H}_{Breit}) \quad (62)$$

where

$$B(i,j) = -\frac{e^2}{2r_{ij}}\left[\alpha_i \cdot \alpha_j + \frac{\alpha_i \cdot r_{ij} \alpha_j \cdot r_{ij}}{r_{ij}^2}\right] \sim mc^2 \alpha^4 Z^4. \quad (63)$$

The Breit interaction energy in He ground state ~ $10^5$ MHz, and in Ne ground state ~ $10^8$ MHz.

The 2p$_{1/2}$-2p$_{3/2}$ fine structure in H atom is 1.095×10$^4$ MHz. Breit interaction cannot be directly observed whereas the fine structure is obtained from spectroscopy Sucher advocated the use of projected Breit operator $\Lambda_+ H_{Breit} \Lambda_+$ to avoid continuum dissolution [2].

**Hyperfine Interaction**

Two-fermion formulation by Chraplyvy [33-34] and subsequent development by Barker and Glover [35] show the hyperfine interaction

$$\hat{H}_{hf} = -\sum_{n}^{\substack{fermion \\ (nucleus)}} \frac{Z_n e^2 \hbar^2 g_e g_n}{4mMc^2} \times \int d^3r \, \psi_+^\dagger(r)\beta_{el}\sigma_{D,el}\psi_+(r)\cdot\beta_n \quad (64)$$

$$\times \left( \frac{\sigma_{D,nu}}{|r-R_n|^3} - 3\frac{(r-R_n)\sigma_{D,nu}\cdot(r-R_n)}{|r-R_n|^5} - \frac{8\pi}{3}\sigma_{D,nu}\delta^3(r-R_n) \right)$$

This may be added to $\hat{H}_{D,ext+}$ following (24) and the QED Hamiltonian $\hat{H}_{QED}$ in (43). The hyperfine splitting is of order $m^2c^2\alpha^4Z^{1-3}/M \sim (m/M)\alpha^2Z^{1-3} \ll \alpha^3Z^3$ hartree, $M$ being the nuclear mass. The hyperfine splitting of H atom is 1420 MHz in ground state, 177 MHz in 2S$_{1/2}$ state, and 59 MHz in 2p$_{1/2}$ and 2p$_{3/2}$ states [30, 32].

Hyperfine structure of Cd$^+$ has been calculated by Li et al. using the relativistic CC [36].

Thus an effective Hamiltonian has been defined,

$$\hat{H}_{QED}^{eff} = \langle \hat{H}_{QED} e^{\hat{T}_{int}^{(2)}} \rangle_{\Psi_{rad}^0} / \langle e^{\hat{T}_{int}^{(2)}} \rangle_{\Psi_{rad}^0} = \left( \hat{H}_{D,ext} + \hat{H}_C + E_{rad}^0 + \hat{H}_{Lamb} + \hat{H}_{Breit,++} + \hat{H}_{hf} \right) \quad (65)$$

where the hyperfine interaction has been added to complement the electronic Breit operator, and following Sucher's suggestion, Breit operator has been considered in the projected form.

## 10. Effects of Matter Cluster

**Negative energy solutions**

Both positive-energy spinors (bound state and continuum solutions) and negative-energy spinors (continuum solutions) together form an orthonormal complete set. A trial square-integrable spinor can be written as a linear combination of the eigenvectors of both positive energy (usually representing bound state components only) and negative energy (spurious solutions). This leads to the possibility of variation collapse [37] and a min-max principle for solving the involved wave equation [23], [25], [38-42].

It is generally taken for granted that the positive-energy Dirac or Dirac-Fock eigenspinors representing bound states form a complete space for normalizable solutions. Though this assumption is wrong, it has been deeply entrenched in quantum chemical calculations.

In this work it is assumed that the Dirac-Fock orbitals have been obtained from the min-max principle discussed in ref. 25.

The negative energy solutions (in practice, the eigenvectors) must be included in the calculation of correlation energy. In field theory, the Coulomb pair operator is obtained as $\hat{H}_C^{Pair} = \hat{H}_C - \hat{H}_C^{np}$.

For a finite basis calculation in Fock space, an approximation to the no-pair Coulomb operator can be obtained by defining $\psi_+$ in terms of positive energy eigenvectors and $\psi_-$ using the spurious eigenvectors of negative energy. The pair operators would be accordingly defined.

**Blocked Pair**

Additional QED correction terms are known to arise from the polarization of vacuum due to the creation and annihilation of virtual electron-positron pairs using the operator. Such energy corrections are mostly blocked because of the exclusion principle. After Pauli blocking, the 1-pair and 2-pair contributions to energy appear as tiny positive corrections of orders $mc^2\alpha^6Z^6$ and $mc^2\alpha^8Z^8$, respectively.

Pair terms do not appear in a relativistic configuration interaction (RCI) calculation that is based on the configurations prepared from only the DF positive-energy eigenvectors (PERCI). The 1-pair (and 2-pair) term(s) appear(s) when the all-energy eigenvectors are considered (AERCI), that is, (spurious) negative energy solutions from the DF are included to obtain de-excitations from the ground state configuration in the RCI. The AERCI corresponds to a many-electron min-max procedure, and the (AERCI – PERCI) energy difference was shown [38-39] to be in excellent agreement with an analytical estimate made by Sucher of the 1-pair energy for helium-like species when $\alpha Z \leq 0.2$ [43].

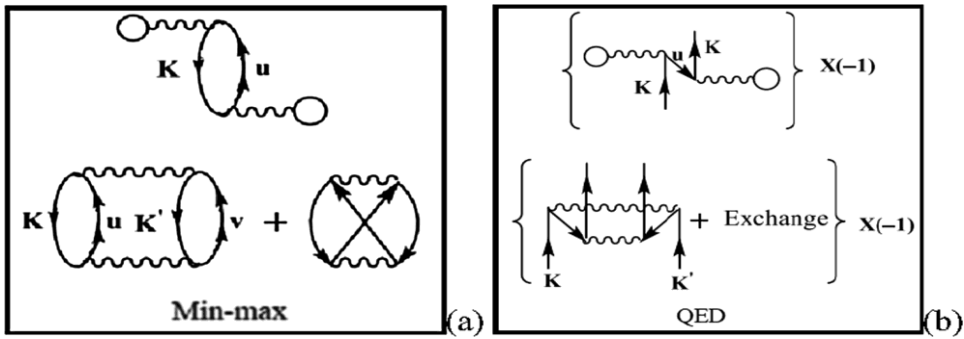

**Figure 1:** (a) (a) Second order energy diagrams representing the admixing of deexcited configurations in MP2 or during AERCI, equivalent to the min-max procedure. A negative-energy electron line (u or v) proceeds upward, and a hole line (**K** or **K′**) runs downward. (b) Second order propagator diagrams with an electron line (**K** or **K′**) proceeding upward and a positron line (u) moving downward.

The vacuum polarization effect on energy is fundamentally a correlation effect, and it can be realized from the cluster operator technique if one considers the more complete Coulomb interaction while exploring the influence of a more detailed matter cluster. However, the pair effects would be prominent for large $\alpha Z$.

Correlation effects are determined from the matter clusters. However, the cluster now includes not only the excitations from the DF ground state configuration to the conventional virtual orbitals but also a mixture of excitations to the virtuals and de-excitations to the spurious levels (indicated by primes) in addition to double deexcitations.

### Correlation Treatment (CS)

Use the matter cluster that has been extended to accommodate the pair operator in

$$\langle \Psi_{rad}^0 | \hat{H}_{QED} - [(E_N^0 + E_{rad}^0) + \langle (\hat{H}_{Breit} + \hat{H}_{Lamb} + \hat{H}_{hf}) \rangle_{\Psi_N^0}] | \Phi_0 \rangle$$
$$= \left( \hat{H}_{QED}^{eff} - [(E_N^0 + E_{rad}^0) + (E_{Breit}^0 + E_{Lamb}^0 + E_{hf}^0)] \right) e^{\hat{T}_{mat}} | \Psi_N^0 \rangle = E_{correl} e^{\hat{T}_{mat}} | \Psi_N^0 \rangle \quad . (66)$$

This leads to a hierarchy of matrix equations which can be solved and $E_{correl}$ can be determined. Examples are shown in Appendix 1 and Appendix 2.

### Summary

<u>Breit</u> interaction was added to (Dirac) Fock space CC as an afterthought. Here it is directly involved, at par with Coulomb interaction, though smaller by an order of $\alpha^2 Z^2$. The coefficients of the exponential cluster $\hat{T}_2$ are determined by it. Breit interaction relies on the inequality $(\hbar c k)^{-1} | E_{DF}^n - E_{DF}^0 | \ll 1$ so that Pair clusters and it do not greatly affect each other. Breit interaction energies can be estimated for both atoms and molecules. Breit interaction energies cannot be directly observed.

The Lamb shift, generally of order $mc^2 \alpha^5 Z^4$, is updated to the many-body level as a natural recourse. Nonzero matrix elements of the charge current can be obtained from two states differing by a single excitation. Hence singly excited intermediate electronic states in $\hat{H}_{Lamb}$ can contribute to correlation energy and coefficients. Lamb shift can be observed for atoms.

The electron-nucleus hyperfine interactions are one-electron effects. In principle they can contribute to the correlation energy and the correlated wave function. These contributions would be more extensive in CCSD, CCSD (T), etc. The hyperfine splitting can be estimated with relative ease. It is detectable from magnetic resonance spectroscopies even for heavier atoms and molecules.

For lighter atoms and molecules, intricate spectroscopic features such as energy ordering of electronic states and the spin-orbit splitting can be observed. Most of the radiative effects remain concealed in a molecule because of extensive rotational and vibrational energy contributions as well as rovibronic and vibronic interactions at a finite temperature.

### General Conclusions

The effective cluster considered in the present work has been

$$\hat{T}_{eff} = \hat{T}_{int}^{(2)} (1 + \hat{T}_1 + \hat{T}_2 + \hat{T}_3 + \hat{T}_4 + ...) + (\hat{T}_1 + \hat{T}_2 + \hat{T}_2^{1-pair} + \hat{T}_2^{2-pair} + \hat{T}_3 + \hat{T}_4 + ...) \quad (67)$$

s.t. $| \Phi_N^0 \rangle = (1 + \hat{T}_{eff}) | \Psi_N^0 \rangle .$ \quad (68)

The tactic has been to calculate an average over the radiation state at first so that the radiative, matter and pair effects become separated,

$$\langle \hat{H}_{QED}\hat{T}_{eff}\rangle_{\Psi^0_{rad}} = E^0_{rad} + \left(\hat{H}_{Lamb} + \hat{H}_{Breit} + \hat{H}_{hf}\right)(1+[\hat{T}_2+\hat{T}_4]) \quad \text{(Radiative)}$$

$$+ \left(\hat{H}_{D,ext} + \hat{H}_{C++}\right)(1+\hat{T}_2+\hat{T}_4) \quad \text{(Matter)} \quad (69)$$

$$+ \hat{H}_C^{Pair}[(\hat{T}_2^{1-pair} + \hat{T}_2^{2-pair})(1+\hat{T}_2+\hat{T}_4) + \hat{T}_4^{pair}] \quad \text{(Pair)}$$

Factor 1 gives the mean field total energy plus the lowest order radiative and pair corrections. Matter clusters contribute to the correlation energy including pair effects.

The simplest example (Example 3) of N′ non-interacting minimal-basis $H_2$ molecules with 2N′ electrons is discussed in Appendix 3.

**Novelties**

QED interactions (Lamb, Breit and hyperfine) are obtained from a single procedure based on the radiative cluster.

Pair energy is determined from an extended matter cluster formalism.

Additional correlation energy can be had from radiative effects and pair terms, while the option for higher order pair energy in high-Z atoms is kept open.

## 11. Open Shell Modifications

The Fock operator is

(DF for closed shell) $\quad F_{DF} = h_{D,ext} + \upsilon_{DF}(r), \quad \upsilon_{DF}(r) = \sum_{u_{m\sigma}}^{occ} [\mathrm{J}_{m\sigma}(r) - \mathrm{K}_{m\sigma}(r)] \quad (70)$

(MCDF or MCSCF for open shell) $F_{MCDF} = h_{D,ext} + \upsilon_{MCDF}(r), \quad F_{MCSCF} = h_{D,ext} + \upsilon_{MCSCF}(r)$

where $\qquad (71)$

$\upsilon_{MCDF}(r) = \sum_{u_{m\sigma}}^{cs}[\mathrm{J}_{m\sigma}(r) - \mathrm{K}_{m\sigma}(r)] + f \sum_{u_{m\sigma}}^{os}[\mathrm{J}_{m\sigma}(r) - \mathrm{K}_{m\sigma}(r)]$ (showing spinor dependence)

$+ 2\alpha_{Roothaan}\left[\sum_c L_c(r) + f\sum_o [L_o(r) - J_o(r)]\right] - \beta_{Roothaan}\left[\sum_c M_c(r) + f\sum_o [M_o(r) - K_o(r)]\right]$

(showing same spatial dependence)

(72)

and

$L_i\phi = f\sum_o \left[\langle\phi_i|J_o|\phi\rangle\phi_i + \langle\phi_i|\phi\rangle J_o\phi_i\right], \quad M_i\phi = f\sum_o \left[\langle\phi_i|K_o|\phi\rangle\phi_i + \langle\phi_i|\phi\rangle K_o\phi_i\right], \quad (73)$

$\alpha_{Roothaan} = (1-a)/(1-f), \quad \beta_{Roothaan} = (1-b)/(1-f), \qquad (74)$

valid for open-shell atoms of low or medium Z, and highly symmetric small molecular radicals. In the general case, one relies on MCSCF where two sets of coefficients are simultaneously varied and the MCSCF potential is determined from the linear combination coefficients of reference determinants.

**Open-Shell Wave Function**

The final matter state is $|\Psi_{electronic,\alpha}\rangle$. Determinants $\{\Phi_\mu\}$ and $\{\Phi_q\}$ span model spaces $M$ and its complement $\tilde{M}$. Consider the projectors $\hat{P} = \sum_{\mu \in M} |\Phi_\mu\rangle\langle\Phi_\mu|$ and $\hat{Q} = \sum_{q \notin M} |\Phi_q\rangle\langle\Phi_q|$.
Then the ground state configuration is written as

$$|\Psi_{electronic,\alpha}^M\rangle = \hat{P}|\Psi_{electronic,\alpha}\rangle = \sum_{\mu \in M} c_\alpha^\mu |\Phi_\mu\rangle \qquad (75)$$

Also,

$$|\Psi_{electronic,\alpha}^{\tilde{M}}\rangle = \hat{Q}|\Psi_{electronic,\alpha}\rangle = \sum_{q \in \tilde{M}} |\Phi_q\rangle\langle\Phi_q|\Psi_{electronic,\alpha}\rangle . \qquad (76)$$

The first intermediately normalized state is $|\tilde{\Psi}_{electronic,\alpha}\rangle|\Psi_{rad}^0\rangle$ where

$$|\tilde{\Psi}_{electronic,\alpha}\rangle = \sum_{\mu \in M} e^{\hat{T}_\alpha^\mu} |\Phi_\mu\rangle c_\alpha^\mu = \sum_{\mu \in M} (1+\hat{T}_{\alpha 1}^\mu + \hat{T}_{\alpha 2}^\mu + \hat{T}_{\alpha 3}^\mu + \hat{T}_{\alpha 4}^\mu + ...)|\Phi_\mu\rangle c_\alpha^\mu \qquad (77)$$

Clusters are neither hermitean nor anti-hermitean. The final intermediately normalized state is

$$|\Psi_{QED,\alpha}\rangle = |\Psi_{electronic,\alpha}\rangle|\Psi_{rad}^0\rangle = \frac{e^{\hat{T}_{1,int}^{(2)}}}{\langle e^{\hat{T}_{1,int}^{(2)}}\rangle_{\Psi_{rad}^0}} |\tilde{\Psi}_{electronic,\alpha}\rangle|\Psi_{rad}^0\rangle$$

$$= \frac{e^{\hat{T}_{1,int}^{(2)}}}{\langle e^{\hat{T}_{1,int}^{(2)}}\rangle_{\Psi_{rad}^0}} \sum_{\mu \in M} e^{\hat{T}_\alpha^\mu} |\Phi_\mu\rangle c_\alpha^\mu |\Psi_{rad}^0\rangle \qquad (78)$$

s.t. $\langle\Psi_{rad}^0|\langle\Psi_{electronic,\alpha}^M|\Psi_{QED,\alpha}\rangle = 1.$ \qquad (79)

**Dynamical correlation**

The matter cluster now includes not only the excitations but also deexcitations

$$\hat{H}_{QED,OS}|\tilde{\Psi}_{electronuc,\alpha}\rangle = (E_\alpha - E_{rad}^0)|\tilde{\Psi}_{electronuc,\alpha}\rangle = E_{electronic,\alpha}|\tilde{\Psi}_{electronuc,\alpha}\rangle, \qquad (80)$$

$$E_{electronic,\alpha} = \left(E_N^0 + E_{Breit}^0 + E_{Lamb}^0 + E_{hf}^0\right) + E_{correl,\alpha} . \qquad (81)$$

The MCDF/MCSCF ground state configuration is $\Psi_N^0 = |\Psi_{electronic,\alpha}^M\rangle$, $E_{correl,\alpha}$ is the correlation energy calculated from MRCC using the QED OS Hamiltonian operator, and $\langle\hat{H}_{int}^{(2)}\rangle_{rad}$ is to be treated at the MCDF/MCSCF level.

The QED open shell CC can be done now in a straight-forward manner. Define

$$H_{\mu\nu}^{eff} = \langle\Phi_\mu|\hat{H}_{QED,OS} e^{\hat{T}_\alpha^\nu}|\Phi_\nu\rangle \text{ with } \langle\Phi_\mu|e^{\hat{T}_\alpha^\nu}|\Phi_\nu\rangle = \delta_{\mu,\nu} \qquad (82)$$

that leads to

$$\sum_{\mu \in M} H_{\nu\mu}^{eff} c_\alpha^\mu = E_{electronic,\alpha} c_\alpha^\nu . \qquad (83)$$

The cluster amplitudes can be explicitly found. Insert the identity

$$\sum_{\nu \in M} e^{\hat{T}^\mu}|\Phi_\nu\rangle\langle\Phi_\nu|e^{-\hat{T}^\mu} + e^{\hat{T}^\mu}Qe^{-\hat{T}^\mu} = 1 \qquad (84)$$

and consider $e^{-\hat{T}_\alpha^\mu} \hat{H}_{QED,OS} e^{\hat{T}_\alpha^\mu} = \bar{H}_\mu$ for each µ and find analogues of Jeziorski-Monkhorst [16] state specific relations. As the number of equations is less than the number of cluster amplitudes, additional sufficiency conditions are needed.

However, a state-specific formulation in Hilbert space with the prescription $\hat{R}_\alpha^\mu |\Phi_\mu\rangle = 0 \ \forall \ \mu$ would stipulate both size-extensivity and size-consistency, thereby bypassing the sufficiency conditions [44]. It is also completely restrictive, resistant to intruders and has hardly any convergence difficulty.

**Explicit Forms**

Matrix elements of effective Hamiltonian are written as

$$H_{\mu\nu}^{eff} = \langle \Phi_\mu | \hat{H}_{QED,OS} | \Phi_\nu \rangle + \langle \Phi_\mu | \hat{H}_{QED,OS} \hat{T}_{\alpha 1}^\nu | \Phi_\nu \rangle + \langle \Phi_\mu | \hat{H}_{QED,OS} (\tfrac{1}{2!}\hat{T}_{\alpha 1}^{\nu 2} + \hat{T}_{\alpha 2}^\nu) | \Phi_\nu \rangle \\ + \langle \Phi_\mu | \hat{H}_{QED,OS} (\tfrac{1}{3!}\hat{T}_{\alpha 1}^{\nu 3} + \tfrac{1}{2!}(\hat{T}_{\alpha 1}^\nu \hat{T}_{\alpha 2}^\nu + \hat{T}_{\alpha 2}^\nu \hat{T}_{\alpha 1}^\nu) + \hat{T}_{\alpha 3}^\nu) | \Phi_\nu \rangle + ... \quad (85)$$

At Order 0: Matrix elements of $\left(\hat{H}_{D,ext} + \hat{H}_C + \langle \hat{H}_{int}^{(2)} \rangle_{rad}\right) + \left(\hat{H}_{Lamb} + \hat{H}_{Breit,++} + \hat{H}_{hf}\right)$

At Order 1: Non-vanishing results from $\left(\hat{H}_{Lamb} + \hat{H}_{Breit,++} + \hat{H}_{hf}\right)$

At Order 2: Coupling with $\hat{H}_C$ (no-pair and pair), $\hat{H}_{Lamb}$ and $\hat{H}_{Breit,++}$ etc.

The explicit Pair Energy is

$$\delta E_{1-pair} = \frac{1}{2mc^2} \sum_{\substack{i<j \\ a,p}} f_i f_j |\langle ap \| ij \rangle|^2 \sim O(mc^2 \alpha^6 Z^6),$$

$$\delta E_{2-pair} = \frac{1}{4mc^2} \sum_{\substack{i<j \\ p'<q'}} f_i f_j |\langle p'q' \| ij \rangle|^2 \sim O(mc^2 \alpha^8 Z^8). \quad (86)$$

The similarity transformed Hamiltonian turns out as

$$\bar{H}_\mu = \hat{H}_{QED,OS} - [\hat{T}_{\alpha 1}^\mu, \hat{H}_{QED,OS}] + \tfrac{1}{2!}[\hat{T}_{\alpha 1}^\mu, [\hat{T}_{\alpha 1}^\mu, \hat{H}_{QED,OS}]] - [\hat{T}_{\alpha 2}^\mu, \hat{H}_{QED,OS}] \\ - \tfrac{1}{3!}[\hat{T}_{\alpha 1}^\mu, [\hat{T}_{\alpha 1}^\mu, [\hat{T}_{\alpha 1}^\mu, \hat{H}_{QED,OS}]]] + \tfrac{1}{2!}[\hat{T}_{\alpha 1}^\mu, [\hat{T}_{\alpha 2}^\mu, \hat{H}_{QED,OS}]] \\ + \tfrac{1}{2!}[\hat{T}_{\alpha 2}^\mu, [\hat{T}_{\alpha 1}^\mu, \hat{H}_{QED,OS}]] - [\hat{T}_{\alpha 3}^\mu, \hat{H}_{QED,OS}] + ... \quad (87)$$

State specific QED-MRCC equations are

$$\langle \Phi_i^a(\mu) | \bar{H}_\mu | \Phi_\mu \rangle c_\alpha^\mu + \sum_{\nu(\neq \mu)} \langle \Phi_i^a(\mu) | e^{-\hat{T}_\alpha^\mu} e^{\hat{T}_\alpha^\nu} | \Phi_\mu \rangle \langle \Phi_\mu | \bar{H}_\nu | \Phi_\nu \rangle c_\alpha^\nu = 0,$$

$$\langle \Phi_{ij}^{ab}(\mu) | \bar{H}_\mu | \Phi_\mu \rangle c_\alpha^\mu + \sum_{\nu(\neq \mu)} \langle \Phi_{ij}^{ab}(\mu) | e^{-\hat{T}_\alpha^\mu} e^{\hat{T}_\alpha^\nu} | \Phi_\mu \rangle \langle \Phi_\mu | \bar{H}_\nu | \Phi_\nu \rangle c_\alpha^\nu = 0, \quad (88)$$

etc. Here too the Breit, Lamb and pair interactions are directly involved in CC, at par with Coulomb interaction, though smaller by order of $\alpha^2 Z^2$, $\alpha^3 Z^2$ and $\alpha^4 Z^4$-to-$\alpha^6 Z^6$, respectively. The coefficients of the exponential cluster $\hat{T}_{\alpha 2}^\mu$ that are mainly responsible for the conventional Coulomb correlation, become slightly modified by the QED interactions.

**Additional Novelty in MRCC**

There is an additional static correlation when the radiation is not isotropic.

For photons in thermal equilibrium,

$$g_{0k\lambda} = \left(1 - e^{-\hbar\omega_k/\tau}\right)^{1/2} \text{ and } g_{Nk\lambda} = g_{0k\lambda} e^{-N\hbar ck/2\tau} \qquad (89)$$

s.t. $\hbar ck/\tau \gg 1 \Rightarrow g_{0k\lambda} = 1 - \tfrac{1}{2}e^{-\hbar ck/\tau}$, $g_{1k\lambda} = e^{-\hbar ck/2\tau} \sim 0$, $g_{2k\lambda} \sim 0$, etc.

Using $\Omega^{-1}\sum_k \to \dfrac{1}{8\pi^3}\int d^3k$, for isotropic radiation

$$\bar{N}_{k\lambda} = \bar{N}_{k\lambda} = \bar{N}_{k\lambda'} \Leftrightarrow g_{Nk\lambda} = g_{Nk\lambda} = g_{Nk\lambda'} \qquad (90)$$

This eventually gives $\langle \Psi^0_{rad} | \hat{H}^{(2)}_{int} | \Psi^0_{rad} \rangle = 0$.

Any possible change in static correlation would occur only when the radiation is not isotropic. Alternatively, it occurs when the atom or the molecule is not freely tumbling but somehow held in an inert matrix that is placed in the path of a beam of radiation. Therefore the requisite information can be gained from a carefully designed matrix isolation type of spectroscopy.

Two examples, one for an atom and the other for a molecule, are discussed in Appendix 4 and Appendix 5, respectively.

## 12. Concluding remarks

Two thought experiments have been suggested for static correlation. It involves an atom (or a molecule) only in certain electronic states, and only when it faces essentially non-isotropic radiation. It would be stimulating to calculate the changes in energy differences and design experiments to observe them.

The nonrelativistic state-specific MRCC is known to give a robust correlation energy as compared to the observed total energy for very low-Z atoms. For those who have the basic relativistic MRCC computer programs, it will be easy to update the programs for computation of QED effects on other properties. For instance, Li et al. have calculated the hyperfine-structure constants and dipole polarizabilities of $Cd^+$ [36].

Higher order QED effects would be visible only for the high-Z systems.

What is required now is to do very accurate calculations implementing the current procedure into atomic structure codes. It is still difficult to treat a complicated open shell system at the nonrelativistic level to the required accuracy where QED effects become important. This remains as a major challenge for the future.

This presentation is a combined summary of the work described in references [4] on closed shells and [5] on open-shell systems, published in 2019 and 2020, respectively, Later, Sunaga and Saue discussed relativistic coupled cluster theory including QED effects where they considered parity violation in chiral molecules [45], and DePrince (himself and with coworkers) calculated cavity-modulated ionization potentials and electron affinities from quantum electrodynamics coupled-cluster theory [46-47].

## Acknowledgment

IIT Bombay is gratefully acknowledged for its library facilities and infrastructures. The author also thanks the organizers of the 18$^{th}$ Theoretical Chemistry Symposium.

## APPENDIX 1. QED-CCD

Write

$$\langle \Psi_{rad}^0 | \hat{H}_{QED} - [(E_N^0 + E_{rad}^0) + \langle (\hat{H}_{Breit} + \hat{H}_{Lamb} + \hat{H}_{hf}) \rangle_{\Psi_N^0}] | \Phi_0 \rangle$$

$$= \left( \hat{H}_{QED}^{eff} - [(E_N^0 + E_{rad}^0) + (E_{Breit}^0 + E_{Lamb}^0 + E_{hf}^0)] \right) e^{\hat{T}_{mat}} | \Psi_N^0 \rangle = E_{correl} e^{\hat{T}_{mat}} | \Psi_N^0 \rangle$$

(A1.1)

and find

$$\sum_{\substack{m<n \\ r<s}} \langle \Psi_N^0 | \hat{H}_{D,ext} + \hat{H}_{C++} + \hat{H}_{Breit} + \hat{H}_{Lamb} |_{mn}^{rs} \rangle C_{mn}^{rs} = E_{correl},$$

$$\sum_{\substack{m<n \\ r,p'}} \langle \Psi_N^0 | \hat{H}_{D,ext} + \hat{H}_C^{Pair} + \hat{H}_{Breit} + \hat{H}_{Lamb} |_{mn}^{rp'} \rangle C_{mn}^{rp'} = \delta E_{1-pair},$$

(A1.2)

$$\sum_{\substack{m<n \\ p'<q'}} \langle \Psi_N^0 | \hat{H}_{D,ext} + \hat{H}_C^{Pair} + \hat{H}_{Breit} + \hat{H}_{Lamb} |_{mn}^{p'q'} \rangle C_{mn}^{p'q'} = \delta E_{2-pair}.$$

Three sets of equations like

$$\langle_{mn}^{rs} | \hat{H}_{D,ext} + \hat{H}_{C++} + \hat{H}_C^{Pair} + \hat{H}_{Breit} + \hat{H}_{Lamb} | \Psi_N^0 \rangle$$

$$+ \sum_{\substack{p<q \\ t<u}} \langle_{mn}^{rs} | \hat{H}_{D,ext} + \hat{H}_{C++} + \hat{H}_C^{Pair} + \hat{H}_{Breit} + \hat{H}_{Lamb} + \hat{H}_{hf} - (E_N^0 + E_{Breit}^0 + E_{Lamb}^0 + E_{hf}^0) |_{pq}^{tu} \rangle C_{pq}^{tu}$$

$$- \sum_{\substack{p<q \\ t<u}} \langle \Psi_N^0 | \hat{H}_{D,ext} + \hat{H}_{C++} + \hat{H}_C^{Pair} + \hat{H}_{Breit} + \hat{H}_{Lamb} |_{pq}^{tu} \rangle \langle C_{mn}^{rs} * C_{pq}^{tu} \rangle$$

$$+ \sum_{\substack{p<q \\ t,u'}} \langle_{mn}^{rs} | \hat{H}_{D,ext} + \hat{H}_{C++} + \hat{H}_C^{Pair} + \hat{H}_{Breit} + \hat{H}_{Lamb} + \hat{H}_{hf} - (E_N^0 + E_{Breit}^0 + E_{Lamb}^0 + E_{hf}^0) |_{pq}^{t\,u'} \rangle C_{pq}^{t\,u'}$$

$$- \sum_{\substack{p<q \\ t,u'}} \langle \Psi_N^0 | \hat{H}_{D,ext} + \hat{H}_{C++} + \hat{H}_C^{Pair} + \hat{H}_{Breit} + \hat{H}_{Lamb} |_{pq}^{t\,u'} \rangle \langle C_{mn}^{rs} * C_{pq}^{t\,u'} \rangle$$

$$+ \sum_{\substack{p<q \\ t'<u'}} \langle_{mn}^{rs} | \hat{H}_{D,ext} + \hat{H}_{C++} + \hat{H}_C^{Pair} + \hat{H}_{Breit} + \hat{H}_{Lamb} + \hat{H}_{hf} - (E_N^0 + E_{Breit}^0 + E_{Lamb}^0 + E_{hf}^0) |_{pq}^{t'\,u'} \rangle C_{pq}^{t'\,u'}$$

$$- \sum_{\substack{p<q \\ t'<u'}} \langle \Psi_N^0 | \hat{H}_{D,ext} + \hat{H}_{C++} + \hat{H}_C^{Pair} + \hat{H}_{Breit} + \hat{H}_{Lamb} |_{pq}^{t'\,u'} \rangle \langle C_{mn}^{rs} * C_{pq}^{t'\,u'} \rangle = 0$$

(A1.3)

can be solved together to obtain the sets of coefficients $\{ C_{mn}^{rs} \}$, $\{ C_{pq}^{t\,u'} \}$ and $\{ C_{pq}^{t'\,u'} \}$.

The CCD pair energies are good estimates from MBPT/2 [2]. Thus

$$C_{mn}^{rp'} = \langle rp' \| mn \rangle / [\varepsilon_m + \varepsilon_n - \varepsilon_r - \varepsilon_{p'}], \quad C_{mn}^{p'q'} = \langle p'q' \| mn \rangle / [\varepsilon_m + \varepsilon_n - \varepsilon_{p'} - \varepsilon_{q'}]. \quad (A1.4)$$

For $\alpha Z \ll 1$, one obtains

$$C_{mn}^{rp'} \simeq \langle rp' \| mn \rangle / 2mc^2, \quad C_{mn}^{p'q'} \simeq \langle p'q' \| mn \rangle / 4mc^2. \quad (A1.5)$$

Thus

$$\delta E_{1-pair} = \frac{1}{2mc^2} \sum_{\substack{m<n \\ r, p'}} |\langle rp' \| mn \rangle|^2 \sim O(mc^2 \alpha^6 Z^6),$$

$$\delta E_{2-pair} = \frac{1}{4mc^2} \sum_{\substack{m<n \\ p'<q'}} |\langle p'q' \| mn \rangle|^2 \sim O(mc^2 \alpha^8 Z^8). \quad (A1.6)$$

## APPENDIX 2. QED-CCSD

The SD cluster is

$$\hat{T}_{mat} = \hat{T}_1 + \hat{T}_2 + \hat{T}_2^{1-pair} + \hat{T}_2^{2-pair} \quad (A2.1)$$

and the corresponding energy equation is

$$\langle \Psi_N^0 | \left( \hat{H}_{QED}^{eff} - [(E_N^0 + E_{rad}^0) + (E_{Breit}^0 + E_{Lamb}^0 + E_{hf}^0)] \right) e^{\hat{T}_{mat}} | \Psi_N^0 \rangle$$
$$= E_{correl} + \delta E_{1-pair} + \delta E_{2-pair} \quad (A2.2)$$

that gives

$$\langle \Psi_N^0 | \left( \hat{H}_{QED}^{eff} - [(E_N^0 + E_{rad}^0) + (E_{Breit}^0 + E_{Lamb}^0 + E_{hf}^0)] \right) (\hat{T}_1 + \hat{T}_2 + \hat{T}_2^{1-pair} + \hat{T}_2^{2-pair} + \tfrac{1}{2}\hat{T}_1^2) | \Psi_N^0 \rangle$$
$$= E_{correl} + \delta E_{1-pair} + \delta E_{2-pair}. \quad (A2.3)$$

The Hamiltonian can be partitioned as

$$\hat{H}_{QED}^{eff} - (E_{rad}^0 + \hat{H}_{Lamb} + \hat{H}_{hf}) = (\hat{H}_{D,ext}^{no-pair} + \hat{H}_C^{no-pair} + \hat{H}_{Breit,++}) + (\hat{H}_{D,ext}^{Pair} + \hat{H}_C^{Pair}). \quad (A2.4)$$

One obtains the CCSD correlation energy

$$E_{correl} = \sum_{m,r}(h_{D,ext})_{rm} C_m^r + \tfrac{1}{4} \sum_{m,n,r,s} (\langle rs \| mn \rangle + \langle rs \| mn \rangle_B) C_{mn}^{rs} + \tfrac{1}{2} \sum_{m,n,r,s} \langle rs \| mn \rangle C_m^r C_n^s \quad (A2.5)$$

and the CCSD energy

$$E_{CCSD} = (E_N^0 + E_{rad}^0) + (E_{Breit}^0 + E_{Lamb}^0 + E_{hf}^0) + (E_{correl} + \delta E_{1-pair} + \delta E_{2-pair}). \quad (A2.6)$$

The amplitude contributions can be determined from

$$\langle_m^r | \left( \hat{H}_{QED}^{eff} - E_{CCSD} \right)(\hat{T}_1 + \hat{T}_2 + \hat{T}_2^{1-pair} + \hat{T}_2^{2-pair} + \tfrac{1}{2}\hat{T}_1^2) | \Psi_N^0 \rangle = 0,$$

$$\langle_{mn}^{rs} | \hat{H}_{QED}^{eff}(1 + \hat{T}_1 + \hat{T}_2 + \hat{T}_2^{1-pair} + \hat{T}_2^{2-pair} + \tfrac{1}{2}\hat{T}_1^2) - E(\hat{T}_1 + \hat{T}_2 + \hat{T}_2^{1-pair} + \hat{T}_2^{2-pair} + \tfrac{1}{2}\hat{T}_1^2) | \Psi_N^0 \rangle = 0. \quad (A2.7)$$

The main application of CCSD is to CCSD(T) where expansion up to third order is considered using triples $\hat{T}_1\hat{T}_2$ and $\hat{T}_1^3$ in the wave function. Like MRCI and CASPT, CC with SD(T) is very good for ground state when a complete basis set extrapolation is utilized. It would also be possible to construct a state specific MRCC including the QED interactions.

# APPENDIX 3. Non-interacting Minimal-Basis H₂ molecules

Each molecule has two sets of doubly degenerate DF 4-component spinor orbitals ($\psi_{1\uparrow}$, $\psi_{1\downarrow}$) and ($\psi_{2\uparrow}$, $\psi_{2\downarrow}$) corresponding to the bonding and antibonding sigma molecular orbitals (LCAS-MS). The bonding spinors are fully occupied in the DF ground state configuration. Thus

- There are only N' doubles $\left|\begin{array}{c} 2_i \bar{2}_i \\ 1_i \bar{1}_i \end{array}\right\rangle$;
- There are equal coefficients $C$ for each double in the expanded matter cluster;
- For pair corrections, N' one-pair doubles of each type $\left|\begin{array}{c} 2_i \bar{1}'_i \\ 1_i \bar{1}_i \end{array}\right\rangle$, $\left|\begin{array}{c} 2_i \bar{2}'_i \\ 1_i \bar{1}_i \end{array}\right\rangle$, $\left|\begin{array}{c} 1'_i \bar{2}_i \\ 1_i \bar{1}_i \end{array}\right\rangle$ and $\left|\begin{array}{c} 2'_i \bar{2}_i \\ 1_i \bar{1}_i \end{array}\right\rangle$, and N' two-pair doubles of each type $\left|\begin{array}{c} 1'_i \bar{1}'_i \\ 1_i \bar{1}_i \end{array}\right\rangle$, $\left|\begin{array}{c} 1'_i \bar{2}'_i \\ 1_i \bar{1}_i \end{array}\right\rangle$, $\left|\begin{array}{c} 2'_i \bar{1}'_i \\ 1_i \bar{1}_i \end{array}\right\rangle$ and $\left|\begin{array}{c} 2'_i \bar{2}'_i \\ 1_i \bar{1}_i \end{array}\right\rangle$ are involved, the primed being negative energy spinors.

There is no intermediate state to connect with DF ground state configuration through the 3-current, and the contribution of $\hat{H}_{Lamb}$ is zero. Hyperfine corrections for two different electron spins cancel each other in the ground state. Correlation energy is determined only from Coulomb and Breit interactions,

$$E_{correl} = N'C(K_{12} + K_{12}^B) = N'\left[\Delta - [\Delta^2 + (K_{12} + K_{12}^B)^2]^{1/2}\right]. \tag{A3.1}$$

As $\alpha Z \ll 1$, influence of pairs on correlation coefficients is neglected, yielding

$$E_{correl}^{DC} = N'C_{DC}K_{12},$$
$$C_{DC} = K_{12}^{-1}\left[\Delta_{DC} - [\Delta_{DC}^2 + K_{12}^2]^{1/2}\right] \tag{A3.2}$$

with

$$\Delta_{DC} = (E_2^* - E_2^0)/2 = (\varepsilon_2 - \varepsilon_1) + [J_{11} + J_{22}]/2 - 2J_{12} + K_{12}, \tag{A3.3}$$

and

$$E_{correl}^{DCB} = N'C_{DCB}(K_{12} + K_{12}^B),$$
$$C_{DCB} = (K_{12} + K_{12}^B)^{-1}\left[\Delta_{DCB} - [\Delta_{DCB}^2 + (K_{12} + K_{12}^B)^2]^{1/2}\right] \tag{A3.4}$$

with

$$\Delta_{DCB} = (E_2^{*DCB} - E_2^{0DCB})/2 = \Delta_{DC} + (J_{22}^B - J_{11}^B)/2. \tag{A3.5}$$

Correlation energy calc. is manifestly size-extensive,

$$E_{correl}^{DC}/N' = \Delta_{DC} - (\Delta_{DC}^2 + K_{12}^2)^{1/2},$$
$$E_{correl}^{DCB}/N' = \left[\Delta_{DCB} - [\Delta_{DCB}^2 + (K_{12} + K_{12}^B)^2]^{1/2}\right]. \tag{A3.6}$$

Contribution of Breit interaction to DF energy is $J_{11}^B$ per diatomic unit whereas its contribution to correlation energy is much less,

$$(E_{correl}^{DCB} - E_{correl}^{DC})/N' =$$
$$-\left[K_{12}^B(2K_{12} + K_{12}^B) - K_{12}^2(J_{22}^B - J_{11}^B)/2\Delta_{DC}\right]\left[1 - (J_{22}^B - J_{11}^B)/2\Delta_{DC}\right]/2\Delta_{DC} + ... \tag{A3.7}$$

DCI energy is not size-extensive, Dirac-Coulomb energy being $N'^{-1}[\Delta_{DC} - (\Delta_{DC}^2 + N'K_{12}^2)^{1/2}]$ per unit. MP2, though size-extensive, has energy $-K_{12}^2/2(\varepsilon_2 - \varepsilon_1)$.

## Appendix 4. Open-shell atomic state

Nitrogen atom in $2p^3$ electronic configuration with 20 determinants, has the ground state configurations $^4S_{3/2} < {^2D_{5/2}} \sim {^2D_{3/2}}$ (splitting 8.7 cm$^{-1}$) $< {^2P_{1/2}} \sim {^2P_{3/2}}$ (splitting 0.4 cm$^{-1}$). The $^2D_{3/2}$ and $^2P_{3/2}$ states are respectively 19224.5 cm$^{-1}$ and 28838.9 cm$^{-1}$ above the ground state [48-49]. These experimental energies must be accounted for by the results of relativistic MCSCF plus dynamical correlation, while considering Breit, Lamb and Hf interactions, and dynamic correlations for each individual term.

As a gedanken experiment, the operator $\langle \hat{H}^{(2)}_{int} \rangle_{rad}$ in (51) can couple $^2D_{5/2}$ with $^2P_{3/2}$ and $^2D_{3/2}$ with $^2P_{1/2}$, each pair of configurations differing in energy by an amount less than 0.05 atomic unit:

$$\langle ^2D | \langle \Psi^0_{rad} | \hat{H}^{(2)}_{int} | \Psi^0_{rad} \rangle | ^2P \rangle = -\frac{e}{\sqrt{\Omega}} \sum_k \left( \frac{2\pi\hbar c}{k} \right)^{1/2} \sum_\lambda \sum_n g_{nk\lambda} \times$$
$$\left[ \sqrt{n+1}\, g_{(n+1)k\lambda} \int d^3r\, \langle ^2D |: \psi^\dagger(r)\, \alpha\psi(r) :| ^2P \rangle . \lambda e^{ik.r} + \right.$$
$$\left. \sqrt{n}\, g_{(n-1)k\lambda} \int d^3r\, \langle ^2D |: \psi^\dagger(r)\, \alpha\psi(r) :| ^2P \rangle . \lambda^* e^{-ik.r} \right] \approx O(\alpha Z).$$
(A4.1)

This will be important in a spectroscopic study of the L=1 and L=2 ground states of atoms frozen in a matrix, with the excited atoms polarized in a specific direction.

## Appendix 5: Open-shell molecular radical

Consider a chain of $N$ conjugated carbon atoms that is a diradical cation, in essence a monoradical with 2 possible radical centers. The pi orbitals $\pi_{(N+1)/2}$ and $\pi_{(N+3)/2}$ must be strongly localized on two radical centers such as nitronylnitroxide, iminoniroxide, verdazyl, oxo-verdazyl, etc. Determinants for the two localized radicals with odd $N$ are

$$\Phi_{1,1/2} = |\pi_1\bar{\pi}_1 \dots \pi_{(N-1)/2}\bar{\pi}_{(N-1)/2}\pi_{(N+1)/2}| \quad \Leftrightarrow \quad E_1$$
$$\Phi_{1,\bar{1}/2} = |\pi_1\bar{\pi}_1 \dots \pi_{(N-1)/2}\bar{\pi}_{(N-1)/2}\bar{\pi}_{(N+1)/2}| \quad \Leftrightarrow \quad E_1$$
(A5.1)

and

$$\Phi_{2,1/2} = |\pi_1\bar{\pi}_1 \dots \pi_{(N-1)/2}\bar{\pi}_{(N-1)/2}\pi_{(N+3)/2}| \quad \Leftrightarrow \quad E_2$$
$$\Phi_{2,\bar{1}/2} = |\pi_1\bar{\pi}_1 \dots \pi_{(N-1)/2}\bar{\pi}_{(N-1)/2}\bar{\pi}_{(N+3)/2}| \quad \Leftrightarrow \quad E_2$$
(A5.2)

When the SOMOs $\pi_{(N+1)/2}$ and $\pi_{(N+3)/2}$ are nondegenerate, each radical has a typical half-filled shell of one electron in one spatial orbital, and there is little static correlation. Even when the SOMOs are degenerate, the electronic configuration is $\pi^1$, fractional occupancy $f$ is ¼, term is $^2\Pi$ with Roothaan coefficients $a = b = 0$ and again, there is no static correlation.

Nevertheless, configurations $\Phi_1$ and $\Phi_2$ are coupled by $\langle \hat{H}^{(2)}_{int} \rangle_{rad}$,

$$\langle \Phi_1 | \langle \Psi^0_{rad} | \hat{H}^{(2)}_{int} | \Psi^0_{rad} \rangle | \Phi_2 \rangle = -\frac{e}{\sqrt{\Omega}} \sum_k \left( \frac{2\pi\hbar c}{k} \right)^{1/2} \sum_\lambda \sum_n g_{nk\lambda} \times$$
$$\left[ \sqrt{n+1}\, g_{(n+1)k\lambda} \int d^3r\, \pi'^\dagger_{(N+1)/2}(r)\, \alpha.\lambda\, \pi''_{(N+3)/2}(r)\, e^{ik.r} + \right.$$
$$\left. \sqrt{n}\, g_{(n-1)k\lambda} \int d^3r\, \pi'^\dagger_{(N+1)/2}(r)\, \alpha.\lambda^* \pi''_{(N+3)/2}(r) e^{-ik.r} \right] \approx O(\alpha Z).$$
(A5.3)

As conjugated organic molecules can be easily held on a surface or into a matrix, or even turned into a polymer, this interaction energy can be non-zero. Then the electronic wave function in the model space is updated as

$$\Psi_\pi^M = c_\pi^1 \Phi_1 + c_\pi^2 \Phi_2, \quad (c_\pi^1)^2 + (c_\pi^2)^2 = 1$$

$$\Rightarrow \Delta E_{static\ correl} \simeq -\frac{|\langle \Phi_1 | \langle \hat{H}_{int}^{(2)} \rangle_{rad} | \Phi_2 \rangle|^2}{E_2 - E_1} \tag{A5.4}$$

Because $c_\pi^2 \neq 0$, the dynamic correlation is also modified.